\documentclass[12pt]{iopart}

\usepackage{graphicx}
\usepackage{iopams}
\usepackage{hyperref}

\begin{document}

\title[NEMS from high-T$_c$ superconducting crystals of BSCCO]{Nanoelectromechanical resonators from high-T$_c$ superconducting crystals of Bi$_2$Sr$_2$Ca$_1$Cu$_2$O$_{8+\delta}$}

\author{Sudhir~Kumar~Sahu$^1$, Jaykumar~Vaidya$^2$, Felix~Schmidt$^3$, Digambar~Jangade$^2$, Arumugam~Thamizhavel$^2$, Gary~Steele$^3$, Mandar~M.~Deshmukh$^2$, Vibhor~Singh$^1$}
\address{$^1$Department of Physics, Indian Institute of Science, Bangalore-560012 (India)}
\address{$^2$Department of condensed matter physics and material sciences, Tata Institute of Fundamental Research, Mumbai - 400005 (India)}
\address{$^3$Kavli Institute of Nanoscience, Delft University of Technology,
	PO Box 5046, 2600 GA, Delft (The Netherlands)}

\vspace{10pt}

\begin{abstract}
In this report, we present nanoelectromechanical resonators fabricated with thin exfoliated crystals of a 
high-T$_c$ cuprate superconductor Bi$_2$Sr$_2$Ca$_1$Cu$_2$O$_{8+\delta}$. 
The mechanical readout is performed by capacitively coupling their motion to a coplanar 
waveguide microwave cavity fabricated with a superconducting alloy of molybdenum-rhenium.
We demonstrate mechanical frequency tunability with external dc-bias voltage, and quality
factors up to $\sim$~36600. Our spectroscopic and time-domain measurements show
that mechanical dissipation in these systems is limited by the contact resistance arising 
from resistive outer layers. The temperature dependence of dissipation indicates
the presence of tunneling states, further suggesting that their intrinsic performance could 
be as good as other two-dimensional atomic crystals such as graphene.
\end{abstract}

%
%
%
\maketitle
%
%

\section{Introduction}

Two-dimensional (2D) atomic crystals host a unique set of electrical, mechanical, 
and optical properties, which are significantly different than their bulk counterparts, 
and are important for their applications towards devices \cite{castro_neto_electronic_2009,
	castellanos-gomez_mechanics_2015,bernardi_optical_2016}. Understanding properties of 2D materials like graphene, NbSe$_2$, MoS$_2$, and black 
phosphorous has provided insight into the mechanical response \cite{bunch_electromechanical_2007,
	chen_performance_2009,singh_probing_2010,sengupta_electromechanical_2010,
	castellanos-gomez_elastic_2012,wang_2016}. These insights include ultra-high mechanical strength, 
complex nonlinearities, anisotropic elastic properties, and coupling with correlated ground 
state like charge density wave \cite{bunch_electromechanical_2007,eichler_nonlinear_2011,
	singh_negative_2016,sengupta_electromechanical_2010,wang_2016}. From the family of 2D materials, 
atomically thin crystals of high-transition temperature (T$_c$) superconductor 
Bi$_2$Sr$_2$Ca$_1$Cu$_2$O$_{8+\delta}$ (BSCCO) were exfoliated in the seminal work by 
Novoselov et al. \cite{novoselov_two-dimensional_2005}, and have been studied to probe their 
phase diagram in few unit cell thick crystals \cite{sandilands_stability_2010,
	wang_micro-raman_2011,huang_reliable_2015}.

High-T$_c$ superconductors host a rich variety of phases, which are important for the 
understanding of the microscopic mechanism of superconductivity in these 
materials \cite{sachdev_colloquium:_2003}. Exfoliable thin superconducting 
crystals provide an avenue to study different quantum phase with in-situ tuning of carrier 
density, revealing nature of the two-dimensional superconductivity 
\cite{liao_superconductorinsulator_2018,zhao_sign_2018}. 
Naturally, exploring the elastic properties of high-T$_c$ superconductors 
with temperature, in the few unit cell limit is interesting for fundamental and applied aspects. 

Apart from their rich electronic properties, thin superconducting crystals 
are attractive for developing cavity-optomechanical systems \cite{aspelmeyer_cavity_2014}. 
Their low mass and hence large quantum zero-point fluctuations make them attractive 
for achieving large coupling strength to electromagnetic fields. 
Recently, there has been an intense interest in exploring materials like graphene, 
NbSe$_2$ etc. to develop optomechanical devices \cite{singh_optomechanical_2014,weber_coupling_2014}, where
high conductivity nature of these materials tends to minimize the resistive dissipation 
for microwave signal \cite{will_high_2017}. 
Moreover, a sensitive detection of elastic properties by cavity optomechanics
could be an interesting technique for investigating phase-transition in these materials, providing additional insights which may not be captured by the conventional measurements of 
thermodynamic quantities \cite{shekhter_bounding_2013, varma_specific_2015}.

Here, we take the first step in this direction by probing an
optomechanical system consisting of multilayer BSCCO membrane 
coupled to a microwave cavity. Our device-design enables addition 
of a dc bias voltage, which is used to tune the resonant 
frequency of mechanical resonator. We measure dissipation 
at ultra-low temperatures in these resonators using spectroscopic 
and time-domain techniques. Based on capacitive-circuit damping model, we quantify the dc-bias dependence of dissipation. These observations suggest that the quality factor of BSCCO is primarily 
limited by the contact resistance arising from the resistive outer layers.

\section{Device fabrication}

\begin{figure*}
\begin{center}
\includegraphics[width = 110 mm]{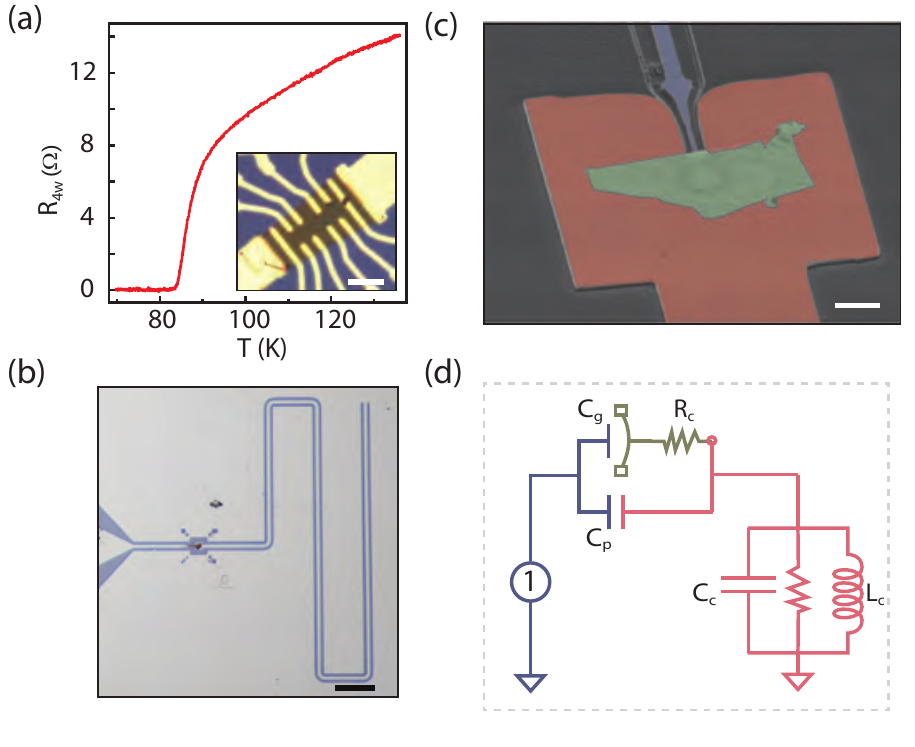}
\end{center}
\caption{(a) Resistance measurements on a flake of BSCCO showing a superconducting transition temperature $T_c$ of 83~K. The inset shows a microscope image of a multilayer BSCCO device used for resistance measurement. The scale bar corresponds to 10~$\mu$m. (b) An image of a single-port quarter-wavelength reflection-cavity in coplanar waveguide geometry fabricated with 300~nm Mo-Re deposited on an intrinsic Si substrate. The scale bar corresponds to 300~$\mu$m. (c) False-color scanning electron microscope image of the coupler region. An almost-circular	drumhead-shaped mechanical resonator can be seen. The scale bar corresponds to 5~$\mu$m. (d) A schematic of the device. The contact resistance between the Mo-Re cavity and BSCCO flake is represented by $R_c$.}\label{fig1}
\end{figure*}

The BSCCO crystals in the form of whiskers were grown using a previously reported technique \cite{jindal_growth_2017}. Due to the sensitivity of BSCCO to the ambient conditions, the grown whiskers were stored at liquid nitrogen temperature to keep them away from moisture, and to suppress the mobility of interstitial oxygen dopants. These whiskers were taken out from liquid nitrogen only at the time of exfoliation. Fig.~\ref{fig1}(a) shows the measurement of 4-probe resistance with temperature for a flake of 40~nm thickness, revealing a superconducting transition temperature of 83~K. All the devices discussed here were fabricated by mechanical exfoliation from whiskers grown in this run. 

To perform readout of the nanomechanical resonators of BSCCO at low temperatures, we capacitively couple their motion to a microwave cavity. Here a thin exfoliated flake from BSCCO crystals is used to form a coupling capacitor to a 
quarter wavelength coplanar waveguide resonator having 50~$\Omega$ impedance and fabricated with an alloy of molybdenum and rhenium (Mo-Re) which has a superconducting transition temperature $T_c \sim $~11~K. Alloys of Mo-Re have been reported to show high-quality factor and low contact resistance in a wide variety of systems \cite{singh_optomechanical_2014,singh_molybdenum-rhenium_2014, amet_supercurrent_2016,kroll_magnetic_2018}. The surface of Mo-Re offers a better adhesion with exfoliated flakes compared to other commonly used superconductors such as aluminum which are prone to surface oxidation. 

To fabricate the microwave cavity, an intrinsic silicon wafer was first extensively cleaned using nitric acid, followed by a 5 min long dip in hydrofluoric acid, and finally a rinse in running DI water. A cleaned silicon wafer was then immediately loaded into the sputter chamber to minimize the oxidation of pristine Si-substrate. A 300~nm thick film of Mo-Re is then deposited using DC-magnetron sputtering. To pattern the Mo-Re film, we prepare a trilayer etch mask, consisting of : 50~$\%$ of diluted-LOR in cyclo-pentanone, 15~nm Al-layer, and 250~nm of PMMA~950~A4 e-beam resist. The top e-beam resist is patterned by lithography. Chlorine plasma is then used to etch the thin Al layer, and subsequently O$_2$ plasma is used to etch the underneath LOR layer. To etch Mo-Re, we have used SF$_6$ plasma. Due to the immunity of Al layer to the SF$_6$ plasma, it acts as a good mask for the patterning of Mo-Re. The etch mask is finally stripped-off using a photo-developer and N-Methyl-2-pyrrolidone. With an additional step of lithography and etching, the Mo-Re film is thinned down near the coupling to the external feedline, and thus forms a part of the capacitive coupler. Fig.~\ref{fig1}(b) shows an optical microscope image of a fabricated quarter-wavelength cavity with Mo-Re on an intrinsic silicon substrate. It is worth to point out here that using the resist stack and lithography steps mentioned above, we consistently measure internal dissipation rates of microwave resonators to be less than
100~kHz on intrinsic Si substrate.

BSCCO flakes are then exfoliated in the ambient environment and transferred to the cavity using a deterministic dry-transfer technique \cite{castellanos-gomez_deterministic_2014}. Fig.~\ref{fig1}(c) shows a zoomed in SEM image of a BSCCO flake in the shape of a drumhead mechanical resonator, forming a coupling capacitor to the cavity. It has a thickness of 54 nm and is suspended by 178 nm over the bottom plate, partially visible through the flake. Fig.~\ref{fig1}(d) shows the schematic of the equivalent lumped-element model. From simulations, we estimated the equivalent cavity parameters $C_p$, $C_c$ and $L_c$ as 6.2~fF, 380~fF and 1.6 nH, respectively. Assuming a parallel-plate capacitor model, the capacitance between the feedline and BSCCO flake was estimated to be 1.8~fF. Using these device parameters, we expect single photon coupling strength to be 0.8~Hz.

\section{Measurements}
\subsection{Cavity characterization}

\begin{figure}
\begin{center}
\includegraphics[width=110 mm]{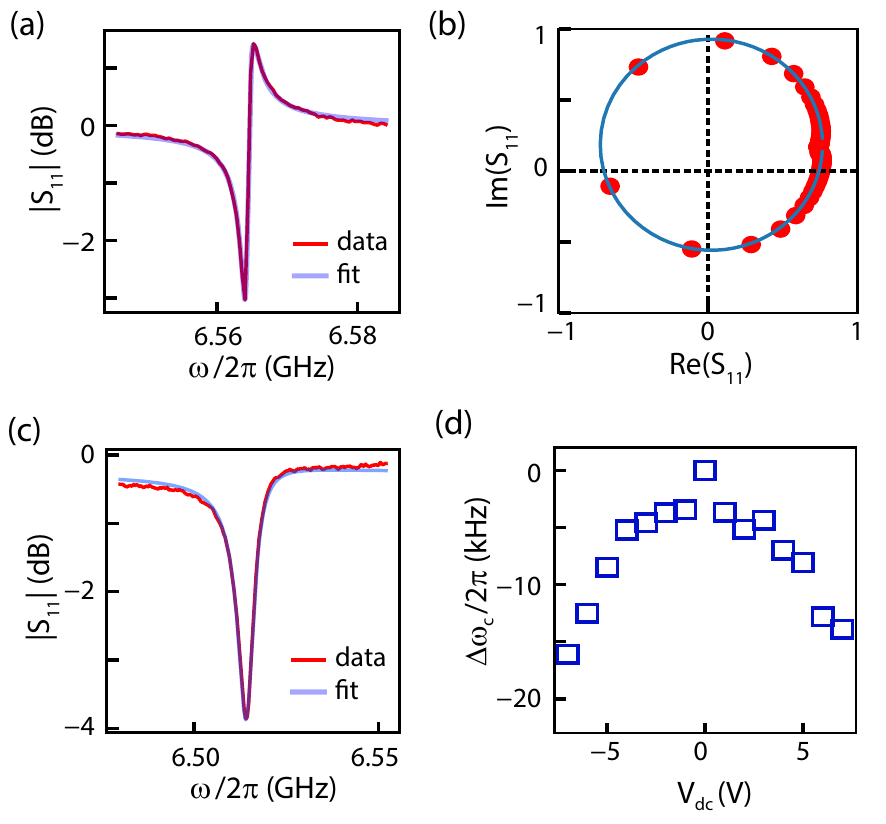}
\end{center}
\caption{(a) Measurement of the reflection coefficient $|S_{11}|$ at 20~mK, along with the fitted curve, yielding internal and external dissipation rates of $2\pi\times$100~kHz and $2\pi\times$1.1~MHz, respectively.(b) Measurement of $S_{11}$ (same as (a)) in polar format, showing the overcoupled nature of the cavity. (c) Measurement of $|S_{11}|$ at 3~K, along with a fitted curve based on the model described in the main text. (d) Measurement of a shift in the cavity frequency $\Delta\omega_c$ with dc voltage applied to the feedline.}\label{fig2}
\end{figure}

For low temperature measurements, the device is placed inside a light-tight sample box machined from OFHC copper, and mounted to the mixing chamber plate of a dilution refrigerator. A sufficiently attenuated input line is used to drive the cavity. Using a circulator, the reflected signal from the cavity is routed towards a HEMT amplifier at the 4K stage. Between the circulator and the device port, we use a broadband bias-tee, which allows us to add a dc voltage to the input feedline of the cavity. At room temperature, we use a vector network analyzer to record the reflection coefficient $|S_{11}|$ of the device. Fig.~\ref{fig2}(a) and (b) show the measurement of $S_{11}$ at 20~mK in the logarithmic and polar format. 
The asymmetry in the resonance curve arises due to the finite isolation of the circulator that is being used to separate input and output signals. This asymmetry can easily be captured by considering a direct leakage of signal from the input port to the output port. The leakage signal and the signal reflected from the cavity interfere with each other to produce a Fano-lineshape in the response. In the polar format, this effect shows up as a resonance circle with its center shifted away from the x-axis. This interference effect can easily be captured by,

\begin{equation}
S_{11}(\omega)=\alpha e^{i\phi}+(1-\alpha)\left(1 - \frac{2\eta}{1+\frac{2i(\omega-\omega_c)}{\kappa}}\right).
\end{equation}

Here $\alpha$ is the isolation of the circulator, $\eta$ is the ratio between the external dissipation rate $\kappa_e$ and the total dissipation rate $\kappa$, $\omega_c$ is the cavity resonant frequency and $\phi$ is the phase-factor arising from the propagation delay. Using this model, we estimated internal and external dissipation rates of $\kappa_i\sim2\pi\times$~101~kHz and $\kappa_e\sim2\pi\times$~1.1~MHz, respectively. The extracted circulator-isolation of 15~dB matches close to the value specified by the manufacturer. As shown in Fig.~\ref{fig2}(c), the validity of this model is further tested by fitting the measurements of $|S_{11}|$ of the cavity at 3~K, reaffirming an external coupling rate of $2\pi\times$~1.1~MHz.

Due to the placement of the mechanical resonator as a part of the quarter-wavelength cavity, a dc bias at the feedline allows us to apply an electrostatic force on the mechanical resonator, thereby changing its mean position. Fig.~\ref{fig2}(d) shows a change in cavity resonance frequency with dc voltage, arising due to the capacitive loading of the cavity-mode. 
We observe a change of $\sim$~15~kHz in the cavity resonant frequency 
with an application of $V_{dc}=~$7~V, corresponding to an estimated displacement of $\sim$~220~pm in the equilibrium position 
of the BSCCO resonator. 

\subsection{Measurement of the mechanical mode}

\begin{figure*}
\begin{center}
\includegraphics[width = 140 mm]{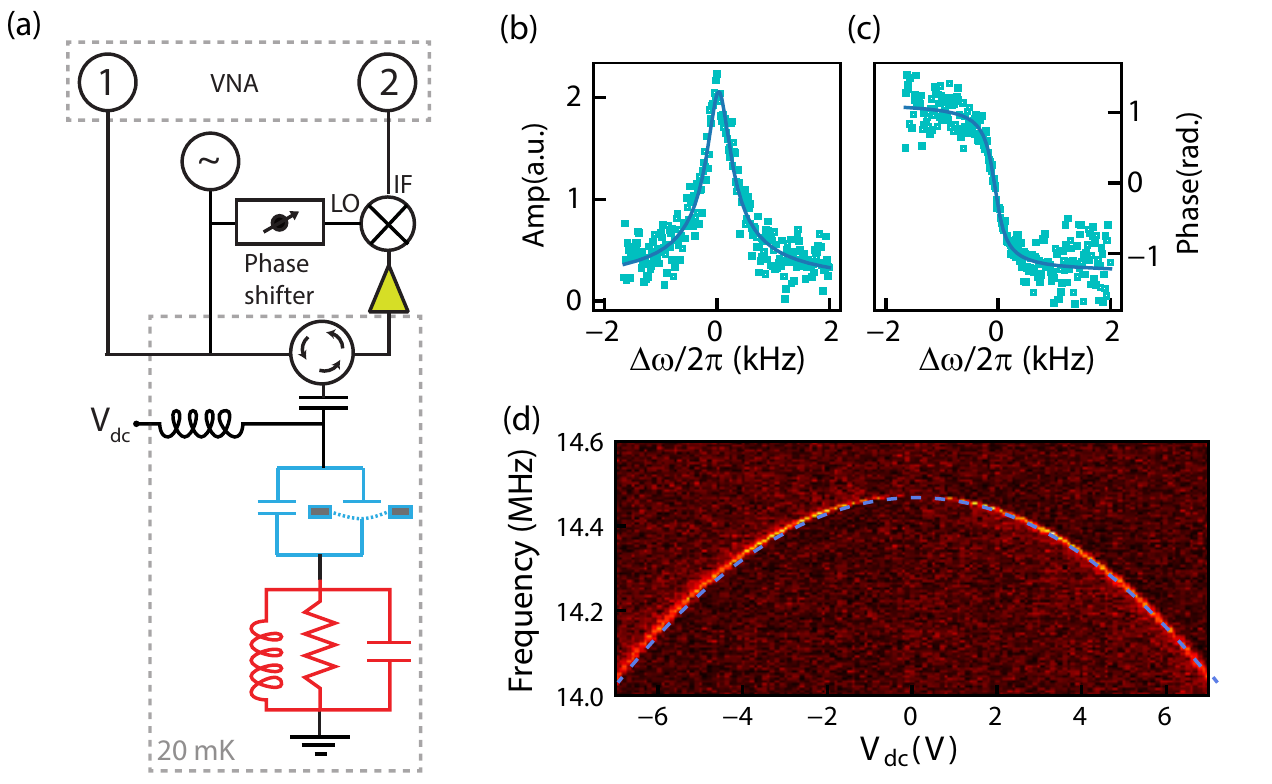}
\end{center}
\caption{(a) A schematic of the homodyne detection setup using a vector network analyzer (VNA). A local oscillator (LO) is used to demodulate the reflected signal from cavity, and the intermediate frequency (IF) signal from an external mixer is sent to receiving port of VNA. (b) and (c) show the amplitude and phase response of the mechanical resonator along with the fitted curve giving a resonant frequency of 14.46~MHz and a mechanical quality-factor $Q_m$ of 36,620. (d) A color plot showing a quadratic tunning of mechanical resonant frequency with dc voltage. The dotted line is the calculated dispersion of mechanical frequency based on geometrical parameters of the device, without any free parameter.}\label{fig3}
\end{figure*}

To characterize the mechanical resonator, we use a homodyne measurement scheme as shown in Fig.~\ref{fig3}(a). A low-frequency RF signal $V_{ac}$ and a dc signal $V_{dc}$ are sent to the feedline to ``electrostatically'' apply a force $C^{'}_gV_{dc}V_{ac}$ on the mechanical resonator, where $C^{'}_g$ is the derivative of flake capacitance with respect to displacement. A microwave tone resonant with cavity frequency is also added to the feedline. The motion of the mechanical resonator phase modulates the reflected signal from the cavity producing two sidebands at $\omega_{c} \pm \omega_{ac}$. The reflected signal is then demodulated using an external mixer at room temperature. The phase of the LO is adjusted so that the mixed down signal is predominantly in the phase quadrature. This way the cavity is used as an interferometer, and phase modulation due to mechanical motion gets recorded. To avoid any nonlinear mixing between the ac and microwave signals in the amplification chain, we use two bandpass filters (2.9~GHz~-~8.7~GHz) before the low temperature amplifier providing $\sim$~80~dB suppression of ac-signal. The ac-drive and the demodulated signals are controlled and recorded by a vector network analyzer, thereby a measurement of $S_{11}$ directly relates to the responsivity of the mechanical resonator.

Fig.~\ref{fig3}(b) and (c) show the mechanical device response measured at $V_{dc}~=~0.3$~V, along with a fitted curve, giving a resonant frequency of 14.46~MHz, and quality factor $Q_m$ of 36620.
Application of the dc voltage enables us to change the equilibrium position of the mechanical resonator, and thus changing its resonant frequency. Fig.~\ref{fig3}(d) shows a color plot of demodulated signal 
as the dc voltage and frequency of ac-drive are varied.
The sharp changes in color represent the resonant frequency of the mechanical mode, showing a parabolic dependence on $V_{dc}$.
It is worth pointing out here that with the ability to add a coherent mechanical drive in our system, a parametric drive can lead to interesting opto-mechanically-induced transparency regimes \cite{zhou_phase_2013, lemonde_enhanced_2016}.

The negative dispersion of mechanical resonant frequency is quite common in electromechanical resonators at low temperatures. It arises from the softening of the mechanical spring constant due to contribution from the electrostatic energy \cite{kozinsky_tuning_2006}. Assuming a parallel plate geometry, for the mechanical capacitor, the dispersion of resonant frequency can be captured by the following equation of motion,

\begin{equation}
m\frac{\partial^2 x}{\partial t^2}+\left(k-\frac{\epsilon_0 \pi r^2V_{dc}^2}{d^3}\right)x=\frac{\epsilon_0  \pi r^2V_{dc}^2}{2d^2} .
\end{equation}

Here $r$ is the radius of the capacitor plate, $d$ is the separation between the two plates, $k$ is the intrinsic spring constant, and $m$ is the total mass of the resonator. It should be noted that due to imperfect clamping, the mode shape could be different from that of an ideal circular drumhead. The choice of the total mass of the resonator leads to an effective amplitude of vibration. The blue dotted line in Fig.\ref{fig3}(d) is the calculated dispersion based on the Eq.(2), using only the device geometry parameters.

\section{Dissipation in mechanical resonators of BSCCO}
\subsection{DC-bias dependence of mechanical response}

\begin{figure*}
\begin{center}
\includegraphics[width = 145 mm]{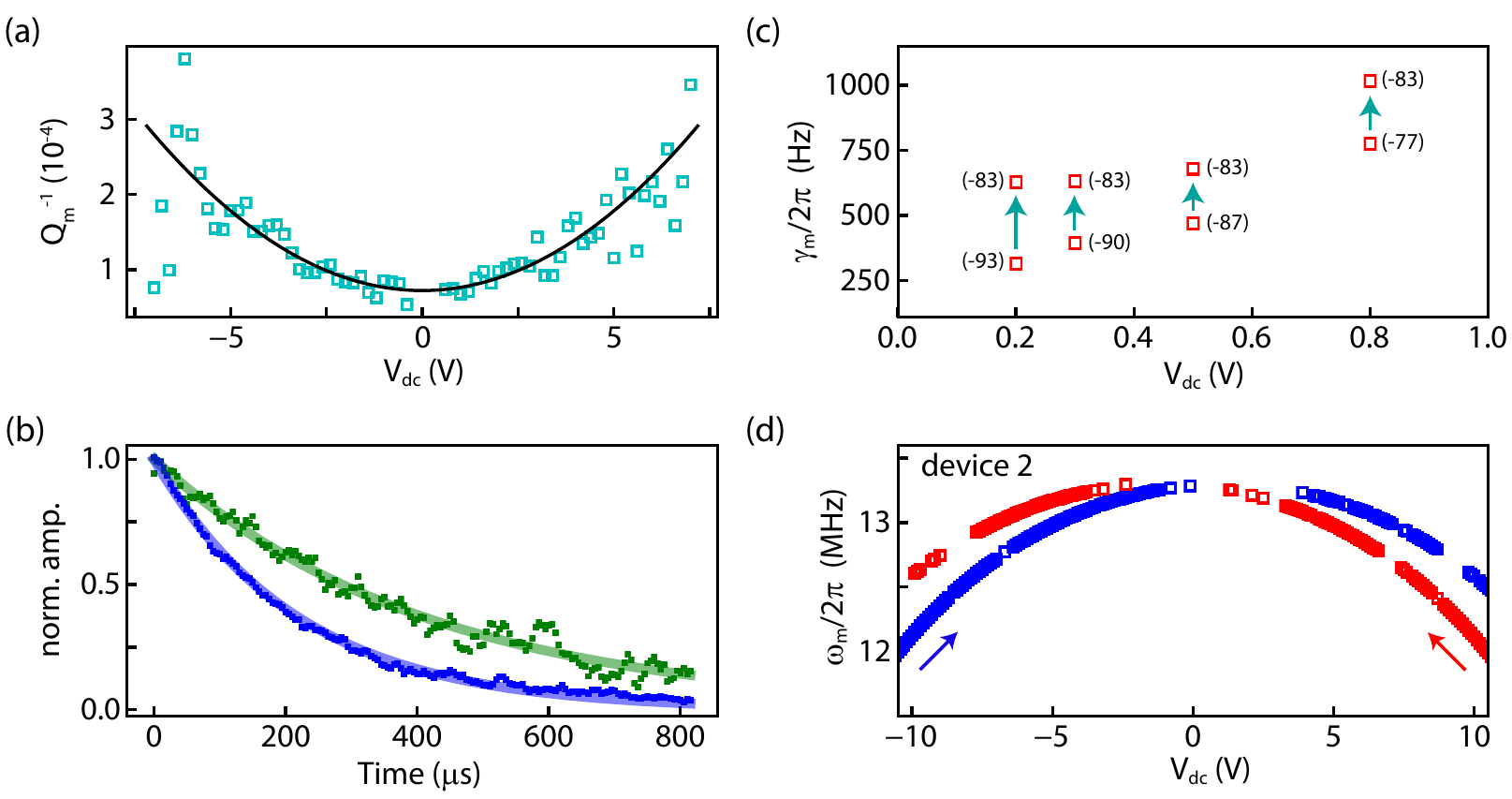}
\end{center}
\caption{(a) Plot of inverse of the mechanical quality factor with $V_{dc}$. The solid line is a fit to the model described in the main text. (b) Time-domain ring-down measurements at $V_{dc}$ = 2~V~(green), and 4~V~(blue). (c) Variation in the mechanical linewidth measured for different injected power in the cavity indicated by the values in dBm inside the parenthesis. (d) Mechanical frequency dispersion with $V_{dc}$ for different sweep directions indicated by arrows.}\label{fig4}
\end{figure*}

After describing the basic measurement scheme, we investigate the dissipation of mechanical mode in this section. In the devices studied here, we observe a strong dependence of quality factor on $V_{dc}$. Fig.~\ref{fig4}(a) shows the plot of inverse of the quality factor with dc voltage. The dc voltage dependent damping can be attributed to the capacitive-circuit damping originating from the finite contact resistance between BSCCO flake and Mo-Re film. For a fixed dc voltage on the feedline, a vibrating capacitor must shuffle charges to the cavity to balance the current. Any finite contact resistance would lead to a power-dissipation, and hence a loss of the mechanical energy. These losses can be worked out as $Q_{c}^{-1}(V_{dc}) = \frac{R_c C_g^{'}V_{dc}^2}{m\omega_m}$ \cite{imboden_dissipation_2014}. The solid line in Fig.~\ref{fig4}(a) plots the total loss-rate of the resonator \textit{i.e.} $Q_m^{-1} = Q_i^{-1} + Q_c^{-1}(V_{dc})$, where $Q_i^{-1}$ is the intrinsic loss rate of the resonator which includes dissipation due to all the other mechanisms.

To rule out broadening of spectroscopic linewidth by resonator dephasing, we performed measurements of mechanical relaxation rate using time domain techniques. This measurement helps in calculating mechanical energy dissipation by removing the sensitivity to first-order frequency noise, and to the mechanical nonlinearities \cite{schneider_observation_2014}. For these measurements, a pulse modulated signal is generated to drive the mechanical resonator, and demodulated signal is recorded on a scope. To improve the signal to noise ratio, ten thousand single shot traces were averaged. Fig.~\ref{fig4}(b) shows the normalized amplitude traces of ring-downs along with fitted curves at two different dc voltages. The extracted relaxation rates match well with the rates calculated from the spectroscopy measurements, thereby ruling out any significant dephasing of the resonator.

Using the capacitive-circuit damping model (solid line in Fig.~\ref{fig4}(a)), we estimate $\sim$~11.5~k$\Omega$ contact resistance between the BSCCO flake and Mo-Re contact. The sensitivity of outer few layers of BSCCO to the ambient conditions is well known, resulting in large contact resistance or even insulating behavior in monolayer samples \cite{novoselov_two-dimensional_2005,sandilands_origin_2014}. 
The contact resistance value reported here corroborates well with the electron-transport devices fabricated without in-situ etch of first few layers of the BSCCO flake. 
Effect of contact resistance can also be seen in the dissipation of mechanical energy with change in the power of cavity driving signal used for readout. Due to high contact resistance, the dissipated energy could locally heat the BSCCO at higher cavity power, thus increasing the mechanical damping. As shown in Fig.~\ref{fig4}(c), the increase in mechanical linewidth for increasing input power supports this hypothesis over a range of gate voltages.

The effect of contact resistance on mechanical modes reflects severely in thinner mechanical resonators. Fig.~\ref{fig4}(d) shows the frequency dispersion from a similar device having 14~nm thick BSCCO mechanical resonator. Hysteresis in resonant frequency is observed with the direction of dc voltage sweep. The hysteresis with respect to sweep direction of dc voltage suggests a larger contact resistance, resulting in a ``slower charging'' of the BSCCO-capacitor, and hence a longer time for the mechanical oscillator to reach an equilibrium position.

\subsection{Temperature dependence of mechanical response}

\begin{figure}
\begin{center}
\includegraphics[width = 82 mm]{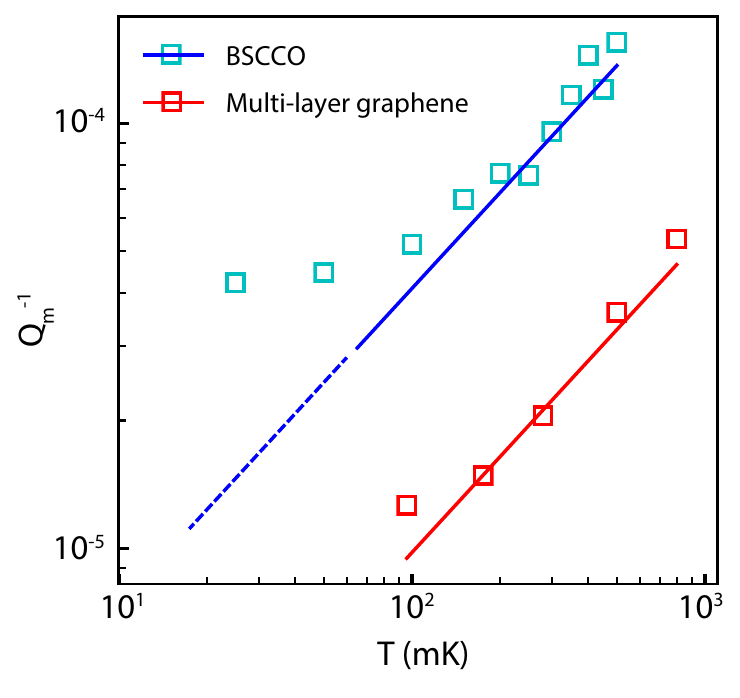}
\end{center}
\caption{Temperature-dependent energy dissipation $(Q_m^{-1})$ of BSCCO mechanical resonator(cyan-squares). The measurements were performed at cavity injected power of -83~dBm and at V$_{dc}$~=~0.6~V. For comparison, similar results from a multilayer graphene resonator are plotted (red-squares). Blue and red solid lines are guides plotted with a temperature exponent of 0.75.}	\label{fig5}
\end{figure}

From the previous discussion on dissipation, it is 
evident that contact resistance at the superconducting interface is degrading the mechanical quality factors of these resonators. It raises an interesting question if the performance of this layered superconductor is limited by resistive heating or by the intrinsic material properties. In the past, resonant ultrasonic spectroscopy techniques have been utilized for the measurements of internal friction and sound velocity in cuprate superconductors, revealing the presence of low energy tunneling states \cite{esquinazi_anomalies_1987, esquinazi_evidence_1988}.

To gain insight into the intrinsic performance of thin BSCCO resonators in the mK range, we record the variations in the mechanical dissipation for different temperatures as shown in Fig.~\ref{fig5}. Such a temperature dependence of dissipation, reflecting a saturation 
at very low temperatures and a power-law dependence above a crossover temperature is quite universal in nano-electro-mechanical resonators of different layered materials \cite{imboden_dissipation_2014}.
Due to large surface to volume ratio in such systems, the enhanced surface defect density  results in low-energy tunneling states \cite{seoanez_surface_2008}.

Depending on the energy distribution of the tunneling states, the mechanical dissipation shows a power-law dependence on 
temperature $(T^\alpha)$, with an exponent that can vary from 0 to 3 below a certain crossover temperature. For example, for 
a broad distribution of energies of tunneling states such as in glasses, dissipation is expected to follow a $T^3$ dependence. 
Crystals with a low density of defects are expected to show a linear dependence and a saturation at lower temperatures \cite{phillips_two-level_1987}. Phenomenological models such as soft-potential model which predicts an exponent of 0.75, and a saturation below a crossover temperature \cite{gil_low-temperature_1993}.
It is also worth pointing out that the molecular dynamics simulation of layered materials, studying friction resulting from 
van der Waals forces between different layers and free edges also predict power-law behavior with exponents varying from 0.3 to 0.9 \cite{kim_multilayer_2009,takamura_energy_2016}.

The observed behavior of dissipation seems to follow the predictions from a soft-potential model. In Fig.~\ref{fig5}, the solid lines are plotted with an exponent of 0.75, serving a guide to the eyes. Similar behavior is observed in multilayer graphene resonator coupled to Mo-Re cavities as shown by red-squares in Fig.~\ref{fig5}. While this observation confirms the presence of tunneling states, the observed exponent seems to suggest the presence of soft localized modes. Recently, a linear temperature dependence of dissipation has been observed in a van der Waals hetero-structure of NbSe$_2$ encapsulated by multilayer graphene attributed to the reduced resistive heating in the mechanical resonator \cite{will_high_2017}. These observations suggest that by improving the electrical contact between BSCCO and Mo-Re surface, the intrinsic performance of BSCCO resonators could be as good as that reported for other multilayer materials. Indeed, by extrapolating the dissipation in BSCCO data, we expect $Q_m >100,000$ at 10~mK temperature.

\section{Conclusions and outlook}

To summarize, we have studied the mechanical performance of the exfoliated thin flakes of a high-T$_c$ superconductor BSCCO in the sub-kelvin temperature range, by coupling their motion to a superconducting cavity. 
These measurements reveal that the mechanical dissipation in these resonators is limited by
the contact resistance between the microwave cavity and BSCCO, resulting from the resistive 
outer layers of the flakes. Temperature dependence of dissipation confirms the presence 
of tunneling states. In future, these issues can be addressed by exfoliation and transfer 
of flakes inside a controlled inert atmosphere, which could provide a way for the applications 
of this material for cavity-optomechanics experiments. 
Moreover, these techniques can be extended to perform measurements near the superconducting 
transition temperature or in presence of the magnetic field to gain insight into the 
microscopics of electron-phonon coupling.

\ack
S.K.S. would like to thank Bindu Gunupudi for providing valuable inputs to the manuscript. This work was supported by STC-ISRO, and the European Union Horizon 2020 research and innovation programme under grant agreement nr. 785219 – GrapheneCore2.
S.K.S. and V.S. acknowledge device fabrication facilities at CeNSE and the Department of Physics IISc-Bangalore funded by DST.
MMD acknowledges Swarnajayanti
Fellowship of DST, Nanomission grant SR/NM/NS-45/2016, and Department of Atomic Energy of Government of India for support.

\end{document}